# Improving Fairness in Speaker Recognition

Gianni Fenu
University of Cagliari
Cagliari, Italy
gianni.fenu@unica.it

Giacomo Medda
University of Cagliari
Cagliari, Italy
g.medda6@studenti.unica.it

Mirko Marras
University of Cagliari
Cagliari, Italy
mirko.marras@unica.it

Giacomo Meloni
University of Cagliari
Cagliari, Italy
g.meloni31@studenti.unica.it

## ABSTRACT

The human voice conveys unique characteristics of an individual, making voice biometrics a key technology for verifying identities in various industries. Despite the impressive progress of speaker recognition systems in terms of accuracy, a number of ethical and legal concerns has been raised, specifically relating to the fairness of such systems. In this paper, we aim to explore the disparity in performance achieved by state-of-the-art deep speaker recognition systems when different groups of individuals characterized by a common sensitive attribute (e.g., gender) are considered. In order to mitigate the unfairness we uncovered by means of an exploratory study, we investigate whether balancing the representation of the different groups of individuals in the training set can lead to a more equal treatment of these demographic groups. Experiments on two state-of-the-art neural architectures and a large-scale public dataset show that models trained with demographically-balanced training sets exhibit a fairer behavior on different groups, while still being accurate. Our study is expected to provide a solid basis for instilling beyond-accuracy objectives (e.g., fairness) in speaker recognition.

## CCS Concepts

• **Computing methodologies~Machine learning~Machine learning approaches~Neural networks** • **Social and professional topics~User characteristics~Gender** • **Social and professional topics~User characteristics~Age**

## Keywords

Speaker Recognition; Speaker Verification; Fairness; Bias; Deep Learning; Discrimination; ResNet; X-Vector.

## 1. INTRODUCTION

*Speaker recognition systems* aim to automatically recognize the identity of an individual from a recording of their voice or speech utterance. These systems have been improved over recent years and have become inexpensive and reliable for person identification and verification [1]. Research in the field of speaker recognition has now spanned several decades and has showed fruitful applications, despite all the different covariates that influence an utterance (e.g., languages, genders, ages, timbres, accents, and background noises). Current successful applications include scanning passengers during border controls, checking identities for bank transactions, forensics analysis, and remote access to computers (e.g., online exams) [2].

When consequential decisions are made about individuals on the basis of the outputs of speaker recognition systems, concerns about discrimination and fairness inevitably arise. Indeed, it may happen that the system's outputs result in decisions systematically biased against individuals with certain protected characteristics like race, gender or age. Underlying patterns of discrimination in the real-world data can be likely picked up in the learning process of the model. This behavior may result in certain groups being unfairly denied access to a platform or being more vulnerable to attackers, with both usability and security issues, respectively. Cognisant of this problem, a timely research paradigm of fair machine learning emerged, attempting to mitigate this unfairness [3,4,5,6]. However, several questions connected to how much unfairness issues affect speaker recognition systems still remain unanswered.

This paper is hence organized around this direction. Our study here aims to raise awareness on the current state of *speaker recognition fairness*, under an identity verification task. First, we manipulated a dataset of voices coming from a range of different demographic groups, identified based on the language (English and Spanish), the gender (male and female), and the age (younger and older than 40 years old). Then, we conducted an exploratory analysis focused on uncovering inequalities exposed by speaker recognition systems to the demographic groups, in terms of both false acceptance and false reject. As a possible countermeasure to the uncovered disparities, we adopted a pre-processing strategy that controls how much each demographic group is represented in the training set. Specifically, the main contribution of this paper is threefold:

- we design a *multi-architecture framework* which makes it possible to train, evaluate, and inspect multiple speaker recognition systems by means of automated pipelines;
- we provide a *fairness-benchmarking* protocol composed by pre-defined training sets and trial recognition pairs for assessing how much speaker recognition systems are fair, based on accuracy disparity among demographic groups;
- we performed an *extensive analysis* of state-of-the-art speaker recognition systems under a setting with eight demographic groups, to investigate on how biases lead to unfairness during the learning process of these systems.

Our experiments allow to better understand how to design speaker recognition systems fairer among protected demographic groups.

This paper is organized as follows. Section 2 describes the related work. Section 3 presents our framework and Section 4 depicts the performed experiments and the more insightful results, showing how balancing the training dataset affects bias mitigation and, by extension, unfairness. Finally, Section 5 summarizes our findings and describes possible future directions.

## 2. RELATED WORK

Machine-learning models have penetrated every aspect of our daily life. Their widespread application has recently raised several social and ethical concerns. One of these concerns is associated with the model's tendency of systematically and repeatedly discriminating an individual or group based on inherent or acquired characteristics (e.g., gender, ethnicity, sexual orientation). Hence, an unfair model is one whose decisions are skewed toward a certain group of people [7]. These ethical aspects have a solid ground in legal frameworks. For instance, explicit mentions are provided in Art. 21 of the EU Charter of Fundamental Rights, Art. 14 of European Convention on Human Rights, where all protected groups are identified, and in Artt. 18-25 of the Treaty on the Functioning of the European Union [8]. Legal definitions of fairness are presented, defining concepts, such as direct and indirect discrimination. Our work finds its core motivation in the latter concept, which is provided in the context of algorithmic decision making. Specifically, indirect discrimination occurs when an apparently neutral rule leads to different outcomes based on a person membership to a protected class [7].

Recently, different studies have evaluated the influence of different factors in the learning process of neural networks and calculated disparities in treatment amongst protected groups [6]. Instilling equality in the outcomes of machine-learning models has gained a lot of attention, especially in the field of face biometrics [9, 10, 11], whereas the literature revealed that there is limited awareness on how unfairness issues affect voice biometrics and, by extension, the resulting speaker recognition systems. Our study here builds upon the preliminary research findings exposed in [5], where the authors uncovered that a deep-learning model exposes different equal error rates among individuals, based on their language, gender, and age. However, their study covered only one type of neural architecture and few demographic groups. Moreover, it has been not considered how changing the representation of the diverse demographic groups in the dataset impacts on the training process as a whole and on the disparities among protected groups. Our work thus aims to provide a better and more general understanding on such primary points.

In general, the literature has showed that one of the main causes of biased results is that models are trained on datasets which do not equally represent the entire population and its demographic groups. Hence, the learning process ends up focusing only on optimizing overall accuracy, performing better only on the groups with high representation in the data. To mitigate this effect, in the context of face biometrics, the authors in [11] have proposed a strategy that aims to use balanced and heterogeneous data to train and evaluate the models. Their results show that training with balanced datasets can partially reduce unfairness. Other countermeasures consist in modifying the objective function used for training the model, to guide learning process into a fairer feature space [9, 10]. This latter strategy is often computationally expensive, and its benefits end up being model dependent. Hence, our preliminary study in this paper aims to mitigate the disparity among protected groups by balancing the dataset for training and testing the models. We believe that such a countermeasure is a primary first step in investigating unfairness in an under-explored domain as speaker recognition systems.

## 3. FRAMEWORK

In this section, we describe the framework we propose in this paper to benchmark and treating unfairness issues in speaker recognition, including the process, its steps, and the main activities (**Fig. 1**). We pre-process the voice-based dataset. Then, we formalize protocols for splitting data in training and testing sets and propose rules to create trial pairs of utterances that simulate real-world recognition process, under identity verification tasks. Models are subsequently trained on the split data. Finally, we define the fairness-benchmark protocol, the disparity metrics, and how to compute them.

### 3.1 Data Organization and Pre-Processing

Our study is carried out on the *FairVoice* dataset proposed in [5], which includes voice data collected under an open source project founded by Mozilla. We selected this dataset since it covers a wide range of demographic groups identified by their language, gender, and age, and the labels which describe such sensitive attributes are available for each individual user. However, not all the languages provided by FairVoice include enough female users in order to set balanced datasets that are sufficiently larger to train state-of-the-art deep speaker recognition systems. Specifically, among the specific-language datasets of English, Spanish, French, German speakers, only the first two languages embrace enough utterances for each demographic group. Thus, we did not consider French and German.

Then, the pre-selected dataset has been filtered by the number of utterances per user. Specifically, only the users who have provided at least five samples were taken into consideration in our analyses. This step is essential because we require to create trial verification pairs with both utterances coming from the same user in order to simulate cases when the authorized user wants to be authenticated. Hence, if a user does not provide a minimum number of utterances, we cannot create enough trial pairs. This filtering step led to a total of 6,321 English speakers and 1,298 Spanish speakers.

In **Table 1**, we report the number of speakers for the gender- and aged-based demographic groups we considered for each language, after applying the above filtering steps. In our preliminary study, we analysed disparities conveyed by speaker recognition systems on four demographic groups per language, based on their gender (female, male) and their age (users younger and older than 40 years old). We selected 40 as a splitting age, since it allows us to better balance the representation of the resulting two age groups, while maintaining a reasonable size of the training dataset. The balancing process in the next steps will be based on the less represented group. For instance, when balancing the Spanish dataset on the number of users per group, we can observe that young females are the less represented, with 180 users. Hence, we filter only 180 users of each group to perfectly balance data across these groups.

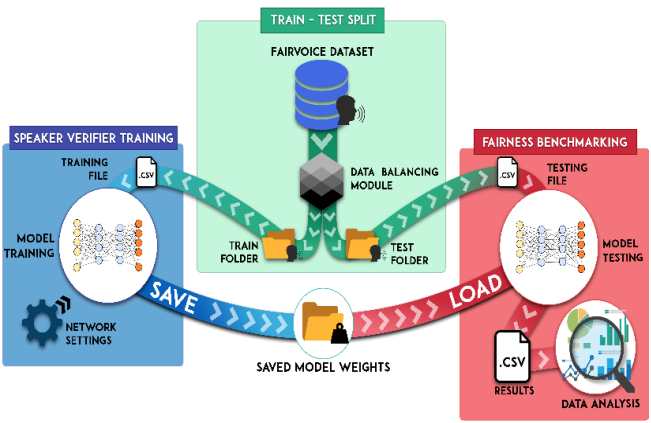

**Figure 1. The fairness-benchmarking framework for speaker recognition systems proposed in this paper.**

**Table 1. Utterances per demographic group embraced into English and Spanish languages considered in our analyses.**

| Language | Group | Speakers |
|---|---|---|
| English | Old females | 425 |
| | Young females | 743 |
| | Old males | 1118 |
| | Young males | 3960 |
| Spanish | Old females | 306 |
| | Young females | 180 |
| | Old males | 376 |
| | Young males | 418 |

## 3.2 Training and Testing Data Preparation

In this step, we are interested in describing the methodologies we used to (i) split data in training and testing sets, (ii) create the trial verification pairs considered to test speaker recognition systems, and (iii) compare their outcomes for the different users' groups.

The utterances belonging to each language included into FairVoice are organized into hierarchical folders, i.e., one folder per language and one subfolder per user in each language folder. Each user's folder contains all the audio samples of that user in a wav format, and each user has a different number of utterances. Each language folder is accompanied by a CSV file that lists, for each user, his/her sensitive attributes (e.g., gender, age, accent) and other statistics.

First, we describe the methodologies we adopted to create testing sets tailored for benchmarking unfairness in speaker recognition. Our methodologies consider selecting a specific number of users for each demographic group and the resulting testing sets contain a range of utterances balanced based on the target sensitive attribute. More precisely, for each language, we selected 100 users, i.e., 25 young females, 25 old females, 25 young males, 25 old males, by randomly sampling them from the corresponding demographic group. We set 100 as the number of testing users in order to include enough users to significantly assess the unfairness of the considered models, and we sampled the same number of users for each group in order to equally cover each group in the testing procedure. Then, for each language, three testing files were created, each with the same amount of trial verification pairs. Each pair includes the references to two utterances in order to simulate a recognition attempt. For each testing file, for each user in the testing set, we created 64 pairs where both the utterances come from the current user (same-user pairs) and 64 utterances where the first utterance comes from the current user and the second one comes from another user (different-user pair). Hence, the three testing files for a given language differ based on the way we selected the *another* user in the different-user pairs, as follows:

- **Test-1**: the *another* user belongs to the same age group of the current user under consideration.
- **Test-2**: the *another* user has the same gender of the current user under consideration.
- **Test-3**: the *another* user is randomly selected.

Finally, we obtain three testing files for each language. We created different testing files to uncover whether the unfairness of a model arises when we compare users belonging to the same demographic group. These files were used to test each model. Note that we repeat these procedures in a multi-fold setting to assess significance.

To organise the balanced training sets, we use the same number of users for each demographic group. For instance, considering the gender, this means having an equal number of male and female users, while considering the age, it means an equal number of old and young users. Combining gender and age, we obtain four groups (i.e., old females, young females, old males, young males) equally distributed in a dataset. In this way models can be trained on more balanced datasets, one of the key strategies used in the state-of-the-art fairness-aware machine learning to reduce unfairness. The audio files used for training included only those of the users who are not considered within the testing set. Based on the learning setup, we chose to train several models, each one with training files having different group-balancing setups, to understand how the training set affects models. Seven training files were created as follows:

- **Train-1** (3 files): for each language configuration, we randomly sampled the same number of *users* for each of the four demographic groups, obtained combining gender and age membership.
- **Train-2** (3 files): for each language configuration, we consider the full dataset of utterances without any type of balancing, i.e., fully unbalanced dataset.
- **Train-3** (1 file): we randomly sampled the same number of *utterances* for each demographic group, obtained combining gender and age membership.

Under each setup, we controlled that the same number of users and of utterances was included across languages, for fair comparison of the results across languages as well. In fact, our study in this paper is interested in evaluating whether the language may be a covariate that leads to unfair performance of a model. This point can promote a more comprehensive understanding of how a speaker recognition model fairly performs in the real world. Thus, we balanced users based on their language, even in the case we trained models with utterances of only a single language.

## 3.3 Speaker Recognition Model Training

Once the list of training utterances was designed by means of the activities described in the previous section, we move to train the speaker recognition models. Specifically, we considered two deep-learning architectures that obtained impressive accuracy results and represent the state-of-the-art in speaker recognition, namely *Thin-ResNet* [12] and *X-Vector* [13]. Note that we decided to select these two architectures due to their different nature. The first one receives audio spectrograms in input and is based on stacked convolutional neural networks, while the second one digests audio filterbanks and is composed by time-delay neural networks. Hence, they give us the opportunity to understand under which circumstances models can lead to a more or less discriminating behavior. Each model was trained for speaker classification in a supervised manner.

By leveraging the training files previously arranged, we were able to train several speaker recognition models under different setups. Finally, 7 model instances were trained for each of the two deep-learning architectures, leading to a total of 14 trained models. For the sake of clarity, we considered the same model parameters and the same training parameters described by the authors of each deep-learning architecture. Even the parameters for acoustic extraction (i.e., spectrogram or filterbank computation) were kept consistent with respect to the original papers [12,13]. Our framework allows to setup the parameters of a training process, e.g., the architecture type (either Thin-ResNet or X-Vector), the number of epochs, the batch size, the learning rate, the optimizer, and so on.

## 3.4 Fairness Benchmarking

In this latter phase, we focused on testing the trained models and analysing how they lead to disparities among demographic groups.

To this end, each trained model was tested with all the three testing files whose creation is described in Section 3.2. Specifically, given a trained model and a testing file, we parsed each trial verification pair, and extracted the acoustic representations (spectrogram or filterbank) from each of the two audio files in the pair. Then, we computed the two corresponding deep speaker embeddings (i.e., the unique numerical representation associated to each audio file) by giving each acoustic representation in input to the model and extracting the embedding from the specific layer of the model. Finally, we get the cosine similarity between the two speaker embeddings. The process was repeated for each trial pair in the testing file, and the results were saved in a CSV that included, in each line, the type of trial pair (0: different-user pair, 1: same-user pair) and the cosine similarity obtained for that trial pair. For each testing results file, we calculated the following set of metrics:

- **Equal Error Rate** (EER): this score represents the error obtained at the threshold where the False Acceptance Rate (FAR) and the False Rejection Rate (FRR) are equal. We computed both the overall EER achieved by a specific model and the EERs achieved by considering only trial pairs coming from a single demographic group.
- **Disparity Score** (DS): this score represents the absolute value of the difference between two EERs, associated with two different demographic groups, e.g., male users' EER and female users' EER or old users' EER and young users' EER. In this way, we can measure disparities in performance based on the EERs. Precisely, a disparity in EERs among groups means that, for that model, can be easier/harder to recognize users within certain groups.
- **False Acceptance Rate (FAR)**: this score measures the chance that the speaker recognition system incorrectly accepts an access attempt by an impostor. FAR results as dividing the number of false acceptances by the number of impostor attempts, and it is associated to the security of the system. The lower the FAR the higher the security.
- **False Rejection Rate**: this score measures the chance the model incorrectly fails to authenticate a legitimate user. FRR is calculated as the ratio between the number of false rejects and the number of genuine attempts, and it is associated to the usability of the system. The lower the FRR, the higher the usability.

The benchmarking phase of all the 14 trained models led to a total of 60 results files.

## 4. EXPERIMENTS

Our study aims to answer to questions on how the decisions of a speaker recognition system, under an identity verification task result in systematically discriminating individuals with protected characteristics like language (assumed here as a proxy of ethnicity), gender, and age. Moreover, we aim to understand how changing the characteristics of the training dataset can reduce the inequality among demographic groups in speaker recognition. Specifically, we will answer to the following four research questions:

- *RQ1: How much is the balancing of the dataset important for mitigating disparities between demographic groups?*
- *RQ2: Is it possible to decrease the model unfairness by leveraging multi-language balanced training datasets?*
- *RQ3: To what extent is the unfairness propagated by the model during the training process?*
- *RQ4: It there any difference in disparity among groups, when different neural architectures are considered?*

Given the large amount of comparisons, we will report only the key and most insightful results and findings that can help to understand how the composition of a training set can be relevant to achieve fair treatments. Note that, in some cases, due to the space constraints, we will not report the results on certain settings, if the uncovered patterns are similar to already-presented ones. The experiments were coded in Python on top of TensorFlow and run on a GPU[1].

### 4.1 RQ1: Influence of Data Balancing

In this first experiment, we trained the two neural architectures using two levels of data balancing in the training set, namely Train-1 and Train-2. Then, we tested each trained model on the three test file, namely Test-1, Test-2, and Test-3. The goal of this experiment is to compare the results obtained based on the metrics we described above and assess whether balancing the number of users per group helps to improve the fairness of the speaker recognition system.

**Table 2** and **Table 3** report the training accuracy, the overall EER, the EERs for each demographic group, and the disparities achieved for gender- and age-based groups by the Thin-ResNet and the X-Vector models, respectively. For the sake of clarity, we discuss the results for each deep-learning architecture independently, leaving a comparison among them to the last research question RQ4. From the results, we can observe the following patterns:

- **X-Vector**: Table 3 shows that slight but not-statistically significant improvements (paired Student t-test, p=0.05) were achieved with the English balanced model (Train-1), when the disparity between ages is considered. Under Spanish, it can be noticed that balancing the number of users per group led to worse results than the unbalanced model.
- **Thin-ResNet**: Table 2 highlights that there is again no significant improvement, both for English and Spanish models, given by the user-balanced dataset. Surprisingly, it seems that the unbalanced dataset gave better results.

These first results point out that a user-balancing strategy is not actually enough to improve the fairness of the pre-trained models. Comparing results across languages, models trained and tested on the Spanish utterances led to a score of disparity significantly lower than the models trained and tested on English utterances, when the same test set is considered. Our experiments expose a situation where, under the English setup, the most discriminated categories were males and young people; under the Spanish setup, a more homogeneous distribution of accuracy across groups was observed.

Even though due to the space constraints we have not reported the corresponding tables, we also examined model's behaviour with respect to the FAR and the FRR achieved for the unbalanced and the user-balanced training set. We observed that X-Vector and the Thin-ResNet prove to be positively influenced by the balancing of the dataset, achieving a significant reduction in the disparity of the FAR and the FRR between males and females and young and old in several cases. Hence, we can conclude that, balancing the dataset made it possible to achieve a fairer treatment of groups with respect to the security and usability of the system (demonstrated by the more similar FAR and FRR), but it did not help the models achieve a more equal performance on recognizing users of some groups.

---

[1] https://mirkomarras.github.io/fair-voice/

**Table 2. RQ1 Thin-ResNet's Disparity Scores (lowest scores under each testing set are highlighted)**

| Language | Train File | Test File | Acc. | EER | EER O | EER Y | EER F | EER M | DS Y/O | DS M/F |
|---|---|---|---|---|---|---|---|---|---|---|
| English | ENGLISH TRAIN 1 | ENGLISH TEST 1 | 87.7% | 6,71 | **5,80** | 7,75 | **4,48** | 8,75 | **1,95** | 4,27 |
|  | ENGLISH TRAIN 2 |  | 71.4% | 7,68 | 4,89 | 10,47 | 5,48 | 8,69 | 5,58 | **3,21** |
|  | ENGLISH TRAIN 1 | ENGLISH TEST 2 | 88.3% | 10,71 | **8,05** | 12,66 | 10,77 | **10,23** | 4,61 | **0,54** |
|  | ENGLISH TRAIN 2 |  | 72.6% | 10,66 | 8,66 | 12,12 | 10,09 | 8,91 | 3,46 | 1,18 |
|  | ENGLISH TRAIN 1 | ENGLISH TEST 3 | 87.7% | 7,24 | **6,17** | 8,05 | 5,61 | 9,02 | 1,88 | 3,41 |
|  | ENGLISH TRAIN 2 |  | 71.4% | 7,91 | 7,75 | 8,25 | **6,89** | 7,66 | **0,50** | **0,77** |
| Spanish | SPANISH TRAIN 1 | SPANISH TEST 1 | 92.2% | 6,56 | 7,17 | **5,88** | 6,02 | 7,11 | 1,29 | 1,09 |
|  | SPANISH TRAIN 2 |  | 89.2% | 6,04 | 6,58 | 5,40 | 6,00 | **5,71** | **1,18** | **0,29** |
|  | SPANISH TRAIN 1 | SPANISH TEST 2 | 92.2% | 8,66 | **7,48** | 9,81 | 9,35 | 7,70 | 2,33 | 1,65 |
|  | SPANISH TRAIN 2 |  | 89.6% | 8,34 | 7,24 | 9,42 | 9,03 | 7,70 | **2,18** | 1,33 |
|  | SPANISH TRAIN 1 | SPANISH TEST 3 | 91.7% | 6,61 | 7,26 | **6,01** | 4,69 | 7,90 | **1,25** | 3,21 |
|  | SPANISH TRAIN 2 |  | 89.6% | 5,89 | 6,95 | 4,54 | **5,21** | 6,44 | 2,41 | **1,23** |

**Table 3. RQ1 X-Vector's Disparity Scores (lowest scores under each testing set are highlighted)**

| Language | Train File | Test File | Acc. | EER | EER O | EER Y | EER F | EER M | DS Y/O | DS M/F |
|---|---|---|---|---|---|---|---|---|---|---|
| English | ENGLISH TRAIN 1 | ENGLISH TEST 1 | 99,1% | 8,17 | **4,69** | 11,66 | **5,64** | 10,50 | 6,97 | 4,86 |
|  | ENGLISH TRAIN 2 |  | 96,2% | 8,20 | 4,70 | 11,69 | 6,06 | 10,38 | 6,99 | **4,32** |
|  | ENGLISH TRAIN 1 | ENGLISH TEST 2 | 96,9% | 10,55 | 7,94 | 13,17 | 8,91 | 12,05 | **5,23** | 3,14 |
|  | ENGLISH TRAIN 2 |  | 98,7% | 10,51 | **7,67** | 13,09 | **8,36** | 10,20 | 5,42 | **1,84** |
|  | ENGLISH TRAIN 1 | ENGLISH TEST 3 | 96,9% | 8,66 | 7,50 | 9,86 | **6,69** | 10,19 | **2,36** | 3,50 |
|  | ENGLISH TRAIN 2 |  | 96,2% | 8,74 | **7,14** | 9,95 | 7,14 | 9,42 | 2,81 | **2,28** |
| Spanish | SPANISH TRAIN 1 | SPANISH TEST 1 | 98,6% | 6,39 | 6,98 | **5,91** | 5,88 | 6,97 | 1,07 | 1,09 |
|  | SPANISH TRAIN 2 |  | 98,6% | 6,39 | 6,70 | 5,75 | **5,98** | 6,52 | **0,95** | **0,54** |
|  | SPANISH TRAIN 1 | SPANISH TEST 2 | 98,6% | 8,46 | **8,06** | 8,63 | 7,81 | 8,75 | **0,57** | 0,94 |
|  | SPANISH TRAIN 2 |  | 97,9% | 8,44 | 7,88 | 8,55 | 8,19 | 8,06 | 0,67 | **0,13** |
|  | SPANISH TRAIN 1 | SPANISH TEST 3 | 98,6% | 6,44 | 8,12 | **4,71** | 4,78 | 8,06 | 3,41 | 3,28 |
|  | SPANISH TRAIN 2 |  | 97,9% | 6,13 | 7,53 | 4,60 | **5,05** | 6,93 | **2,93** | **1,88** |

### 4.2 RQ2: Balanced Multi-Language Training

In this second experiment, we aim to understand whether using a larger dataset that combines both Spanish and English speakers, we can better mitigate disparities between demographic groups. To this end, we merge the training files corresponding to the two languages (e.g., English Train-1 and Spanish Train-1 were fused to obtain a multi-language Train-1 file). Then, each trained model was tested on all the six testing files, three per language, separately.

**Table 4** and **Table 5** show the training accuracy, the overall EER, the EERs for each demographic group, and the disparities under the above setup. As previously done, we present and discuss our results separately for each neural architecture:

- **X-Vector**: Table 4 shows that the variations in disparity between the various tests on the models is minimal. However, both under the Spanish and English languages, distinguishing users from the same age (Test-1) is more challenging than distinguishing those belonging to the same gender group (Test-2). It can be also observed that, for the young-old disparity scores, the best results are reported on the models trained on balanced datasets. Concerning the male-female disparity scores, no clear pattern was highlighted, except that the Train-3 setting performed worse than the others in the balanced test for male-female users (Test-2).

- **Thin-ResNet**: Table 5 reports that there is little evidence of an improvement in terms of disparity by means of the balanced dataset (Train-1/3). Overall, the disparity scores are higher than those seen for the X-Vector architecture. The fairest results in terms of balance are mostly reported in the disparity score between male-female users.

Combining the findings on the first two research questions, it can be observed that there is not enough evidence in order to confirm significant improvements in the disparity score due to the balance of the dataset, even when a model is trained on the same amount of utterances per demographic group. Overall, it is important to note that our findings uncover that these exists less disparity in the Spanish language than in the English language. Considering the protected groups were most discriminated against, we note that the English language exhibits a greater discrimination against males than females and against young than elderly users. On the Spanish language, there is no particular trend of disparities against a specific demographic group according to the tests. This leads us to conclude that, in English, the disparities are more marked and systematically affect the same protected groups, while the Spanish language do not show a systematic behavior against a specific group.

**Table 4. RQ2 X-Vector's Disparity Scores (lowest scores under each testing set are highlighted)**

| Train File | Test File | Acc. | EER | EER O | EER Y | EER F | EER M | DS Y/O | DS M/F |
|---|---|---|---|---|---|---|---|---|---|
| ENGLISH-SPANISH TRAIN 1 | ENGLISH TEST 1 | 99,2% | 7,67 | **4,50** | 10,50 | **5,67** | 9,09 | **6,00** | 3,42 |
| ENGLISH-SPANISH TRAIN 2 | ENGLISH TEST 1 | 98,6% | 7,70 | **4,53** | 10,67 | **5,30** | 9,33 | 6,14 | 4,03 |
| ENGLISH-SPANISH TRAIN 3 | ENGLISH TEST 1 | 98,0% | 7,36 | **4,09** | 10,62 | **5,81** | 9,03 | 6,53 | **3,22** |
| ENGLISH-SPANISH TRAIN 1 | ENGLISH TEST 2 | 99,2% | 9,44 | **6,97** | 11,91 | **8,31** | 9,91 | **4,94** | 1,60 |
| ENGLISH-SPANISH TRAIN 2 | ENGLISH TEST 2 | 98,6% | 9,42 | **6,34** | 12,50 | **8,30** | 9,83 | 6,16 | **1,53** |
| ENGLISH-SPANISH TRAIN 3 | ENGLISH TEST 2 | 98,0% | 9,55 | **6,17** | 12,91 | **8,70** | 10,36 | 6,74 | 1,66 |
| ENGLISH-SPANISH TRAIN 1 | ENGLISH TEST 3 | 99,2% | 8,35 | **6,55** | 9,48 | **6,16** | 9,48 | 2,93 | 3,32 |
| ENGLISH-SPANISH TRAIN 2 | ENGLISH TEST 3 | 98,6% | 7,96 | **6,59** | 9,14 | **6,11** | 8,81 | 2,55 | **2,70** |
| ENGLISH-SPANISH TRAIN 3 | ENGLISH TEST 3 | 98,0% | 8,08 | **6,64** | 9,06 | **6,06** | 9,48 | **2,42** | 3,42 |
| ENGLISH-SPANISH TRAIN 1 | SPANISH TEST 1 | 98,2% | 5,91 | 5,97 | **5,62** | **5,45** | 5,81 | **0,35** | 0,36 |
| ENGLISH-SPANISH TRAIN 2 | SPANISH TEST 1 | 98,6% | 5,50 | **5,11** | 6,02 | **5,19** | 5,62 | 0,91 | 0,43 |
| ENGLISH-SPANISH TRAIN 3 | SPANISH TEST 1 | 98,7% | 5,51 | 5,69 | **5,26** | 5,44 | **5,31** | 0,43 | **0,13** |
| ENGLISH-SPANISH TRAIN 1 | SPANISH TEST 2 | 99,2% | 8,18 | **7,70** | 8,52 | **7,63** | 7,94 | 0,82 | **0,31** |
| ENGLISH-SPANISH TRAIN 2 | SPANISH TEST 2 | 97,8% | 8,28 | **7,65** | 8,54 | 8,05 | **7,52** | 0,89 | 0,53 |
| ENGLISH-SPANISH TRAIN 3 | SPANISH TEST 2 | 99,3% | 7,85 | 7,93 | **7,59** | **7,46** | 8,61 | **0,34** | 1,15 |
| ENGLISH-SPANISH TRAIN 1 | SPANISH TEST 3 | 99,2% | 5,80 | 7,21 | **4,00** | **4,14** | 7,15 | 3,21 | 3,01 |
| ENGLISH-SPANISH TRAIN 2 | SPANISH TEST 3 | 97,8% | 5,61 | 7,29 | **3,79** | **3,94** | 6,31 | 3,50 | 2,37 |
| ENGLISH-SPANISH TRAIN 3 | SPANISH TEST 3 | 98,0% | 5,59 | 7,21 | **4,00** | **4,19** | 6,52 | 3,21 | **2,33** |

**Table 5. RQ2 Thin-ResNet's Disparity Scores (lowest scores under each testing set are highlighted)**

| Train File | Test File | Acc. | EER | EER O | EER Y | EER F | EER M | DS Y/O | DS M/F |
|---|---|---|---|---|---|---|---|---|---|
| ENGLISH-SPANISH TRAIN 1 | ENGLISH TEST 1 | 92,0% | 6,66 | **5,23** | 7,55 | **4,53** | 8,19 | 2,32 | 3,66 |
| ENGLISH-SPANISH TRAIN 2 | ENGLISH TEST 1 | 91,2% | 7,27 | **5,44** | 8,70 | **5,56** | 7,52 | 3,26 | **1,96** |
| ENGLISH-SPANISH TRAIN 3 | ENGLISH TEST 1 | 86,4% | 7,11 | **6,03** | 7,23 | **5,08** | 8,86 | **1,20** | 3,78 |
| ENGLISH-SPANISH TRAIN 1 | ENGLISH TEST 2 | 92,0% | 9,22 | **7,17** | 11,27 | **7,98** | 9,69 | 4,10 | 1,71 |
| ENGLISH-SPANISH TRAIN 2 | ENGLISH TEST 2 | 90,4% | 9,84 | **8,42** | 11,27 | 9,22 | **7,94** | 2,85 | 1,28 |
| ENGLISH-SPANISH TRAIN 3 | ENGLISH TEST 2 | 87,0% | 9,21 | **7,59** | 10,83 | 9,08 | **8,98** | 3,24 | **0,10** |
| ENGLISH-SPANISH TRAIN 1 | ENGLISH TEST 3 | 91,2% | 7,02 | **5,61** | 8,12 | **6,09** | 7,80 | 2,51 | 1,71 |
| ENGLISH-SPANISH TRAIN 2 | ENGLISH TEST 3 | 91,1% | 7,46 | **6,83** | 8,28 | **6,53** | 7,97 | **1,45** | **1,44** |
| ENGLISH-SPANISH TRAIN 3 | ENGLISH TEST 3 | 87,0% | 7,05 | **5,91** | 8,20 | **5,31** | 7,72 | 2,29 | 2,41 |
| ENGLISH-SPANISH TRAIN 1 | SPANISH TEST 1 | 92,0% | 5,76 | 5,84 | **5,66** | 6,13 | **5,23** | **0,18** | 0,90 |
| ENGLISH-SPANISH TRAIN 2 | SPANISH TEST 1 | 91,1% | 5,23 | 5,34 | **5,10** | **4,78** | 5,12 | 0,24 | 0,34 |
| ENGLISH-SPANISH TRAIN 3 | SPANISH TEST 1 | 86,4% | 6,67 | 7,34 | **5,91** | 6,72 | 6,47 | 1,43 | **0,25** |
| ENGLISH-SPANISH TRAIN 1 | SPANISH TEST 2 | 92,0% | 8,67 | **7,04** | 11,05 | 10,78 | **6,52** | 4,01 | 4,26 |
| ENGLISH-SPANISH TRAIN 2 | SPANISH TEST 2 | 90,4% | 8,34 | **5,76** | 11,25 | 10,33 | **5,97** | 5,49 | 4,36 |
| ENGLISH-SPANISH TRAIN 3 | SPANISH TEST 2 | 87,0% | 9,21 | **7,52** | 11,62 | 10,81 | **7,64** | 4,10 | 3,17 |
| ENGLISH-SPANISH TRAIN 1 | SPANISH TEST 3 | 92,0% | 5,88 | 6,72 | **4,97** | **4,54** | 6,66 | 1,75 | 2,12 |
| ENGLISH-SPANISH TRAIN 2 | SPANISH TEST 3 | 91,1% | 5,36 | 5,55 | **5,13** | **4,54** | 5,75 | **0,42** | 1,21 |
| ENGLISH-SPANISH TRAIN 3 | SPANISH TEST 3 | 85,8% | 6,30 | 6,90 | **5,60** | **5,68** | 7,10 | 1,30 | 1,42 |

To answer to our question, the use of a multi-language and balanced training dataset led to improvements in the decrease of the disparity between sensitive categories in more than half of the experiments compared to the single-language trained models. However, this behavior can depend on the diversity of the dataset or on the greater number of samples within the dataset. Our results also uncover that the evaluation protocol has a key impact on assessing disparities.

Even under this research question, we analysed both FAR and FRR of all multi-language models, without reporting the results due to space constraints. It was observed that both X-Vector and Thin-ResNet were positively influenced by the multi-language balanced training set. In fact, the fully balanced model (Train-3) achieved the lowest disparity in FAR among demographic groups in the Spanish language. Similar patterns were observed for FRR. Interestingly, X-Vector achieved good fairness only in young-old disparity, while Thin-ResNet performed well also on the male-female disparity. To conclude, the multi-language balanced datasets encouraged a mitigation of the disparities on the FARs and FRRs.

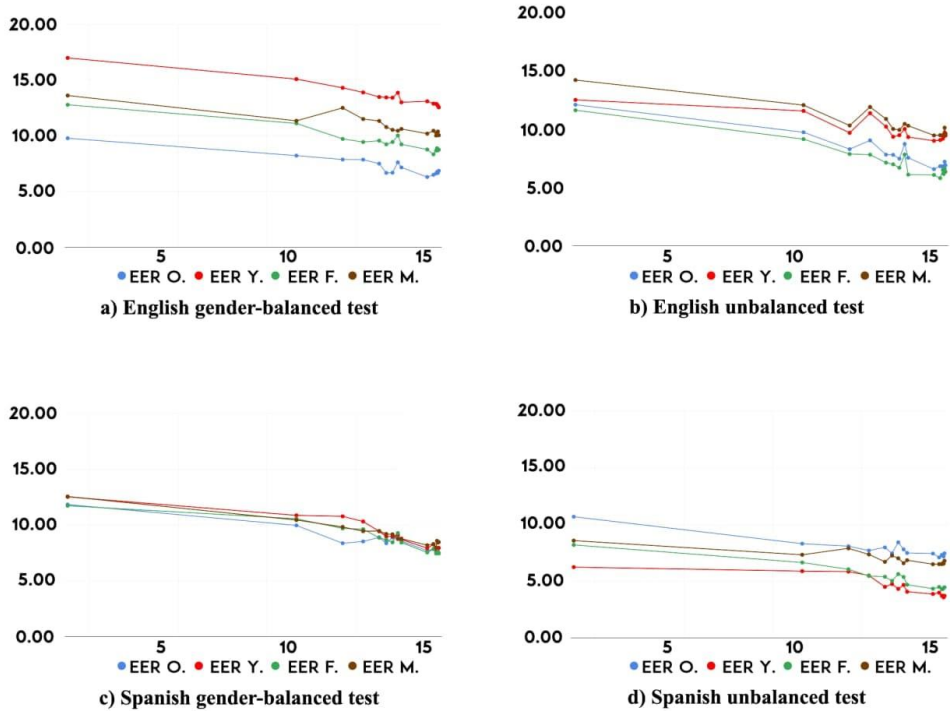

**Figure 2. X-Vector EER variation over epochs**

### 4.3 RQ3: Disparity Propagation during Training

In this experiment, we aim to understand if and how the disparity in treatment among demographic groups is propagated during the training of a speaker recognition model. For the sake of clarity, we focus our discussion on the multi-language balanced model (Train-3), described in the previous section. Similar patterns were obtained for the other model instances. Specifically, we are interested in studying how the disparity score evolves during training, so we saved a copy of the model after each training epoch and computed the metric scores it achieves on the various testing files. This allows us to analyse in more detail the behaviour of the two architectures and the differences between the tested languages. Due to the space constraints, we will focus our attention on two testing setups: the gender-balanced test (Test-2) and the unbalanced test (Test-3).

**Figure 2** and **Figure 3** present the EERs achieved by the models after each training epoch, for each demographic group, under the X-Vector and Thin-ResNet architectures.

- **X-Vector**: it can be observed that, epoch by epoch, both for the English language and the Spanish language, the young-old disparity score in the gender balanced tests (Test-2) decreases compared to the early epochs (**Figure 2a** and **2c**). Clear patterns appear on the results from the unbalanced test (Test-3), where the disparities increase with the progress of the epochs, becoming larger with respect to the early epochs (**Figure 2b** and **2d**).
- **Thin-ResNet**: in these experiments, we have noticed that balancing the tests does not produce a change of trend in unfairness during training. In the English language, there is a very low initial disparity, which tends then to increase during the epochs, even in the young-old balanced test (**Figure 3a** and **3b**). In Spanish, epoch by epoch, the disparities between ages increase compared with the initial value, while for unfairness across gender groups, there is a constant decrease during epochs (**Figure 3c** and **3d**).

To sum up, under the Spanish setup, models have the least disparity between groups compared with English.

### 4.4 RQ4: Comparison across Architectures

Based on the experiments previously performed, we aim to uncover whether the two different neural architectures, X-Vector and Thin-ResNet, show a consistently different behavior. This allows us to understand whether one architecture is more inclined to unfairness.

Considering the disparity scores collected during the experiments, it can be observed that Thin-ResNet exhibits a fairer behavior than X-Vector, though Thin-ResNet has a slower convergence in the learning process than the X-vector. Furthermore, the experiments presented in Section 4.4 show how X-Vector has a more stable performance in EER for each demographic group compared to Thin-ResNet. By comparing the unfairness behaviour resulting from the two architectures on the balanced testing files, X-Vector was more sensitive to variations in the testing set balance, and the lower disparity is registered for balanced training sets, showing large improvements compared to what observed for Thin-ResNet. It follows that X-Vector requires that the testing set is balanced to achieve fair results. Conversely, Thin-ResNet is fairer as more training samples are fed into it, no matter of the demographic group.

With both architectures, training-balanced models led to similar FARs/FRRs among groups, meaning that different demographic groups experience comparable security and usability level in the underlying platform. In Spanish, X-Vector presented the lowest male-female disparity, while Thin-ResNet achieved a lower young-old disparity score. In English, Thin-ResNet is the architecture that presents the fairest behavior, especially in the fully balanced model.

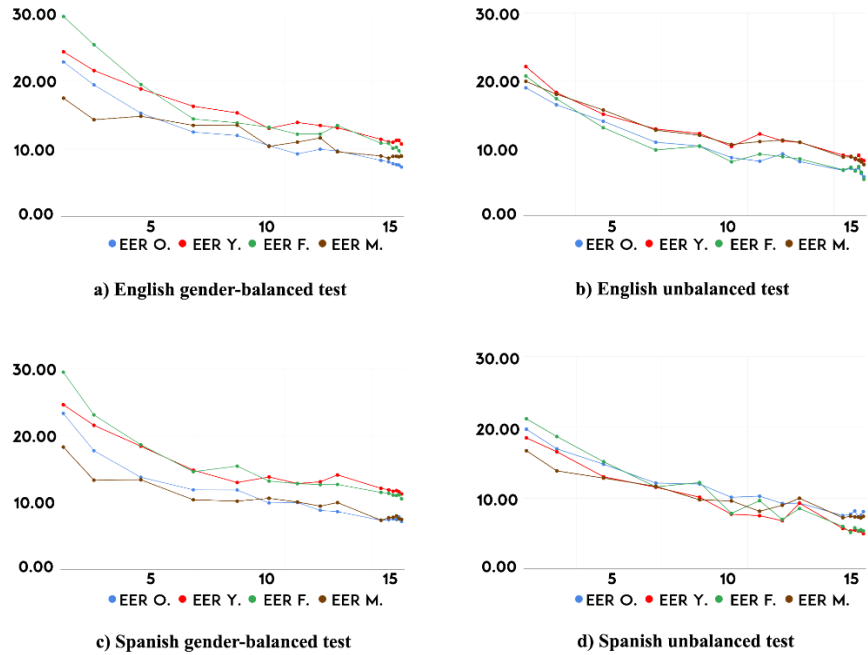

**Figure 3. Thin-ResNet EER variation over epochs**

## 5. CONCLUSIONS

In this paper, we studied to what extent state-of-the-art speaker recognition models systematically expose unfair decisions across demographic groups. Then, we investigated whether it is possible to mitigate their unfair behavior by controlling the representation of demographic groups in the training set. Based on our results:

- Larger and more demographically balanced datasets help to decrease the disparity in EER among protected groups.
- Balancing the training makes it possible to reduce the differences in FARs and FRRs among protected groups, otherwise present under unbalanced training conditions.
- On average, compared to X-Vector, Thin-ResNet leads to a lower disparity among groups and is less sensitive to changes in the balance level of the training set.
- In the English language, models tend to systematically discriminate certain demographic groups, while spurious disparities were noted in the Spanish language.

Future works will enrich our analyses with more languages and demographic groups. Moreover, we will devise novel architectures and optimization approaches to reduce the uncovered unfairness.